\documentclass[twocolumn,prb,preprintnumbers,amsmath,superscriptaddress,amssymb]{revtex4-1}
\usepackage[section]{placeins}
\usepackage{graphicx}
\usepackage{amsmath,amssymb}
\usepackage{color}
\usepackage{multirow}
\usepackage[latin1]{inputenc}
\usepackage[T1]{fontenc}
\usepackage[english]{babel}
\usepackage{amsmath,amssymb,amsthm}
\usepackage{hyperref}
\usepackage{booktabs}
\usepackage{bbm}
\usepackage{url}
\usepackage{nicefrac}
\usepackage{multirow} 
\usepackage{graphicx}
\usepackage{array}
\usepackage{times,txfonts}
\newcommand{\av}[1]{\left\langle #1 \right\rangle}

\newcommand{\qo}[1]{``#1''}

\usepackage{array}
\renewcommand{\epsilon}{\varepsilon}
\renewcommand{\phi}{\varphi}

\newcolumntype{L}[1]{>{\raggedright\let\newline\\\arraybackslash\hspace{0pt}}m{#1}}
\newcolumntype{C}[1]{>{\centering\let\newline\\\arraybackslash\hspace{0pt}}m{#1}}
\newcolumntype{R}[1]{>{\raggedleft\let\newline\\\arraybackslash\hspace{0pt}}m{#1}}
\usepackage{sistyle}
\usepackage{soul}
\definecolor{lightblue}{RGB}{185,210,248}
\sethlcolor{lightblue}

\begin{document}
\title{Observation of Nonclassical Photon Statistics in Single-Bubble Sonoluminescence}
\author{Mohammadreza Rezaee}
\affiliation{Department of Physics, University of Ottawa, 25 Templeton, Ottawa, Ontario, K1N 6N5 Canada}
\author{Yingwen Zhang}
\affiliation{Department of Physics, University of Ottawa, 25 Templeton, Ottawa, Ontario, K1N 6N5 Canada}
\affiliation{National Research Council, 100 Sussex Dr, Ottawa, ON K1A 0R6, Canada}
%%%
\author{James L. Harden}
\affiliation{Department of Physics, University of Ottawa, 25 Templeton, Ottawa, Ontario, K1N 6N5 Canada}
%%%%
\author{Ebrahim Karimi}
\affiliation{Department of Physics, University of Ottawa, 25 Templeton, Ottawa, Ontario, K1N 6N5 Canada}
\affiliation{National Research Council, 100 Sussex Dr, Ottawa, ON K1A 0R6, Canada}
\maketitle
\textbf{A cavitation bubble inside a liquid, under a specific set of conditions, can get trapped in an antinode of the ultrasonically driven standing wave and periodically emits visible photons~\cite{gaitan:90,gaitan1990frontiers}. This conversion of sound to light phenomenon, known as sonoluminescence, can be seen with unaided eyes and occurs in multi- or single-bubble regimes. The sonoluminescence radiation spectrum analysis attributes a temperature of about 6,000 Kelvin to the bubble~\cite{flint:91,brenner:02,xu:10} -- close to the sun's surface temperature. The bubble dynamics, thermodynamics, and other physicochemical properties are well-explored and studied over the past decades~\cite{walton:84, prosperetti:88, putterman:00}. Notwithstanding that several theories such as the thermal (black-body radiation) model~\cite{jarman:60,finch:63,roy:94}, shock wave theory~\cite{wu:94}, phase transition radiation~\cite{tatartchenko:17a,tatartchenko:17b}, and vacuum quantum electrodynamics~\cite{schwinger:92, eberlein:96} -- were proposed, the underlying physics of sonoluminescence has hitherto remained a puzzle. Here, we experimentally investigated the photon number statistics of the emitted photons from single-bubble sonoluminescence (SBSL), employing two distinctly different techniques, i.e., measuring multiphoton correlations and photon number (using a photon-number resolving detector). Our findings show that the emitted photons from SBSL possess sub-Poissonian statistics, indicating the nonclassical nature of the SBSL. Our observation of sub-Poissonian statistics of emitted photons may help to explain the physics of SBSL. In addition, considering the importance of nonclassical light sources in quantum technologies, this discovery would be an exciting milestone in paving the route towards a bright quantum photonics source.}
%231

The dynamics and oscillation of an ultrasonically driven, micron-sized single spherical bubble in a liquid medium kept in thermal equilibrium and ambient pressure conditions can be explained by the Rayleigh-Plesset equation~\cite{rayleigh:17, plesset:48}. Other formalisms have been proposed to describe bubble dynamics in different regimes~\cite{prosperetti:88, flynn:75,keller:80}. In particular, the physics governing the bubble dynamics changes dramatically when the amplitude of the acoustic driving wave reaches a certain threshold. Above this threshold, a bubble trapped in an antinode of a periodically driven standing wave at ultrasound frequency undergoes an abrupt contraction, resulting in extreme temperature and pressure rise and simultaneous light emission. This nonlinear regime, defined by the sound wave's pressure amplitude and the gas concentration, is different for single-bubble sonoluminescence~\cite{gaitan:92} (SBSL) and multi-bubble sonoluminescence~\cite{marinesco:33} (MBSL) -- in this work, we only investigated the SBSL. For SBSL, the radius of a stably trapped air bubble in degassed water, when it emits light, oscillates between 5 and 50 microns, a three order of magnitude change in the bubble volume~\cite{barber:92}. When the bubble periodically collapses to its minimum radius, at the resonant frequency ($\approx30$ kHz), a bright light flash is emitted, containing about $10^6$ photons ~\cite{brenner:02} that propagate in all directions ($4\pi$-steradian). Interestingly, although the time between SBSL triggering states is several tens of microseconds, the emitted pulse duration is in the range of $40-350$ picoseconds depending on the gas concentration and the driving pressure amplitude~\cite{gompf:97,hiller:98,pecha:98}. The spectrum of SBSL light has been recorded and analysed for different bubble regimes. Viewed as black-body radiation, the SBSL emission suggests that the temperature inside the bubble is close to 6,000~K. Notwithstanding existing theories, the physics of phonon-to-photon conversion at the heart of the sonoluminescence effect has remained a puzzle for decades~\cite{maddox:93}. These theoretical models are either based on classical electrodynamics of light generated during bubble collapse (e.g., due to thermally-induced radiation~\cite{vuong1999shock,flannigan2005plasma}, Bremsstrahlung radiation~\cite{moss1997understanding, frommhold1998electron} or radiation from recombination of electrons and ions~\cite{walton1984sonoluminescence}), or on quantum electrodynamics (e.g., radiation due to the quantum nature of the vacuum state~\cite{schwinger1994casimir,eberlein1996theory,milton1997casimir,brevik:99,musha2011another}).
In this study, we employ quantum optics and novel instrument approaches in the hope to elucidate the underlying nature of the light emitted from the collapsing bubble trapped in an acoustic standing wave. We measured and analysed the photon statistics of SBSL using two different experimental approaches: (1) an intensified time-stamping camera to measure high-order temporal correlations among photons emitted from SBSL, and (2) a photon number resolving camera to count the SBSL photons, and consequently their statistics, directly. Both approaches show sub-Poissonian photon statistics indicating the underlying mechanism of SBSL to be nonclassical in nature. Enthralling, measurements based on the second-order correlation function (Hanbury-Brown and Twiss) have been proposed to determine the bubble's physical features~\cite{trentalange:96}, e.g., the light-emitting region size, two-photon correlation measurements as a test for thermality~\cite{belgiorno:00} and the pulse width~\cite{gompf:97}, but not to experimentally explore the fundamental nature of SBSL. 
%487
%
\begin{figure*}[t]
	\centering
	\includegraphics[width=2\columnwidth]{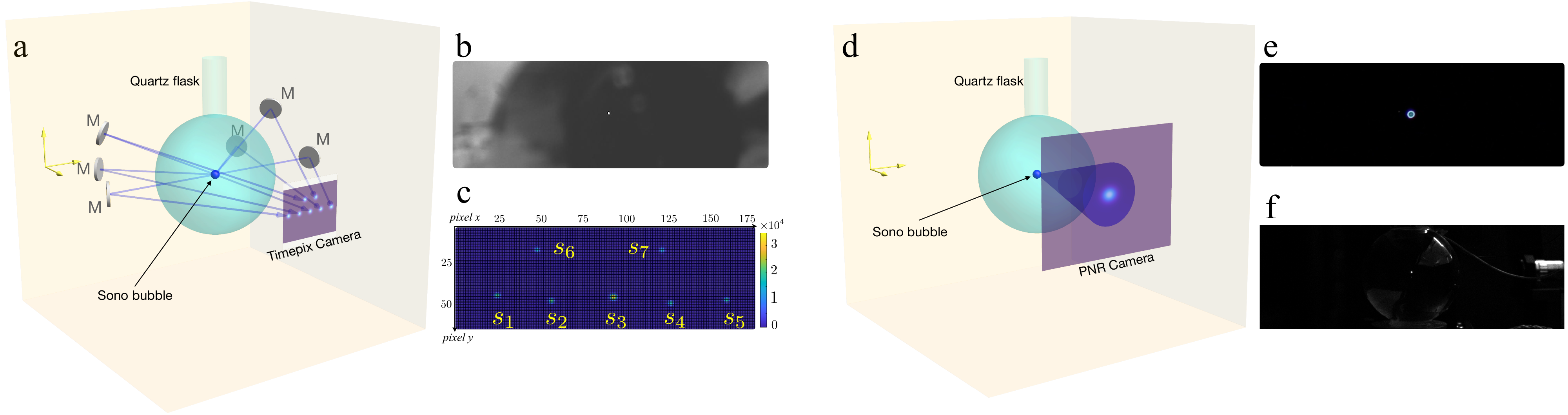}
	\caption{{\bf Experimental setup to measure multiphoton correlation functions from single-bubble sonoluminescence.} {\bf (a)} Schematic of the experimental setup to generate and image single-bubble sonoluminescence onto a time-tagging camera (TPX3CAM) that is made single- photon sensitive with an attached image intensifier (Photonis Cricket with Hi-QE Red photocathode). Six mirrors are used to reflect the emitted photons from the single-bubble, which propagate in different directions onto different regions of the time-tagging camera. {\bf (b)} shows the single-bubble sonoluminescence light that is visible in the round flask. Photo is taken by a monochrome industrial camera with CMOS Pregius sensor using a zoom lens (8\,mm EFL, f/1.4). {\bf (c)} The raw image was taken of photons from the SBSL detected by the camera from seven different directions, ($s_1, s_2, s_3, s_4, s_5, s_6, s_7$), including direct imaging of the source spot at $s_3$, with an exposure time of 22\,s. The propagation length for all directions, excluding the central $s_3$, is about 300\,mm. {\bf (d)} Schematic of the experimental setup for photon number statistics measurement. {\bf (e)} Zoomed-in image of the SBSL taken using a commercial camera. {\bf (f)} The raw image of sonoluminescence setup with SBSL visible in the middle of the flask taken by the ORCA-Quest camera with an exposure time of 10\,s (See Methods for details of both experiments).} 
	\label{fig:fig1}
\end{figure*}

Figure~\ref{fig:fig1} shows the schematic of our experimental setups. Inset (a) is the schematic of the experiments performed using a photon arrival time-tagging camera detector (TPX3CAM)~\cite{nomerotski:19} and (b) is the photon number resolving experiment performed using the ORCA-Quest camera from Hamamatsu. A single air bubble is trapped in degassed water using a standing ultrasonic wave created by a single piezoelectric transducer. The piezo element is attached to the bottom of a $100$ ml spherical quartz flask. See section~\ref{SI:S1} of Supplementary Information (S.I.) for more details on the experimental setup. The air bubble is trapped when the resonance frequency of the water-filled flask is in tune with the driving ultrasound frequency, which is $28.06$ kHz in this case. 

\begin{figure*}[t]
	\centering
	\includegraphics[width=2\columnwidth]{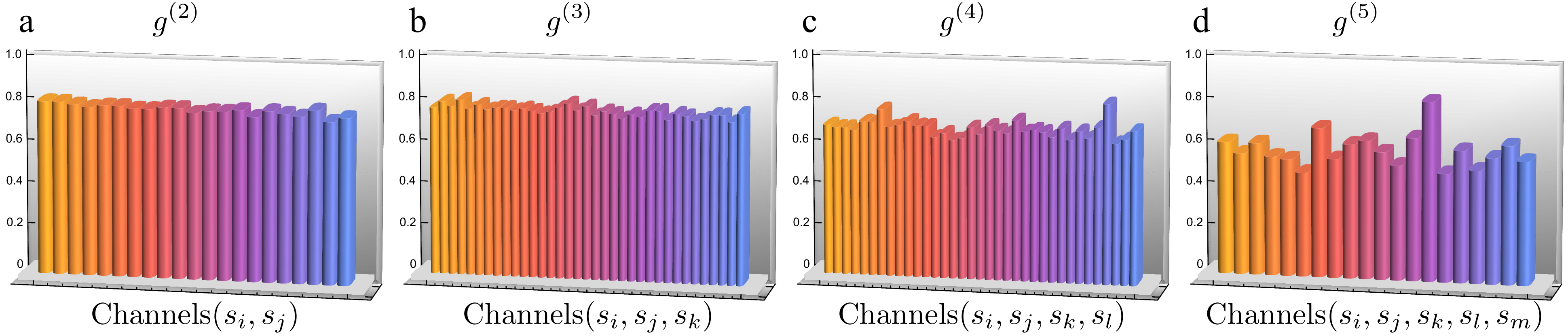}
	\caption{\textbf{Summary of all calculated $g^{(n)}$ for all possible permutations of the seven channels (e.g., ($s_1$,$s_2$); ($s_1$,$s_3$); ($s_1$,$s_3$,$s_4$); etc.) for multi-fold coincidences.} For instance, the first column shows the second-order correlation functions, $g^{(2)}$, for all the possible permutations of channels organized in $n$-fold for $n=2-5$ combinations. The total data acquisition time is 22 seconds, with a $25$~ns gate width. Two-fold column includes all the data points for all the possible permutations, including two channels. The same logic is valid for all other columns. Only events within a 10-pixel radius around the centre of each bright spot (see Fig.~\ref{fig:fig1}-{\bf c}) were taken into account in the data analysis and a constant background of $\simeq4.3$ photons per second per pixel was measured and was subtracted in all calculations.}
	\label{fig:fig2}
\end{figure*}

To explore the photon statistics through temporal correlations between SBSL photons, emitted photons from different directions are collected and analysed as depicted in Fig.~\ref{fig:fig1}-(\textbf{a}). As a first test, photons collected from two different directions, reflected from one of the mirrors ($s_4$) and the source ($s_3$), direct imaging of the SBSL light, were coupled into two multimode optical fibres, prior to being detected by single-photon (avalanche photodiode) detectors, the setup is not shown. Across many runs, a second-order correlation $g^{(2)}$ between 0.8 to 0.9 was  consistently measured, while that measured for a multimode thermal light source and a LED-light were always consistent with 1 (which is expected as the coherence time of classical light sources are much less than the time resolution of the detector) -- see \ref{SI:S2} of the S.I.. Henceforth, we performed a more comprehensive set of tests, where emitted photons from seven different directions were collected and analysed simultaneously. Collected photons are imaged onto a time-tagging camera, which independently offers an effective 7 nanosecond time-stamping resolution on each of its $256\times256$ pixels. Figure~\ref{fig:fig1}-(\textbf{c}) shows a raw image of the photons from SBSL taken by the camera accumulated over 22\,s. The photons from seven different directions are detected as seven bright spots on the camera, denoted by $s_i$, with $i=1-7$. $s_3$ is the direct image of the SBSL captured by the camera, while the other six spots $(s_1,s_2,s_4,s_5,s_6,s_7)$, are SBSL photons reflected by mirrors towards the camera from six different directions. The average photon number detected per pixel for the bright SBSL spots is below 0.03 photon per pulse (or 840 photons per second), this eliminates the possibility of oversaturating the camera chip with more than 1 photon per pulse per pixel resulting in inaccurate photon statistics and, at the same time, guarantees adequate photon statistics analysis can be performed. Temporal correlations of up to $g^{(5)}$  among all seven spots were measured, histograms showing the measured $g^{(n)}$ between different spots can be seen in Fig.~\ref{fig:fig2}. We can see the measured $g^{(n)}$ is consistently below $1$, and decreases with increasing $n$ from $g^{(2)}\simeq0.82$ down to $g^{(5)}\simeq0.6$. Note that for classical and coherent light sources $g^{(n)}$ is expected to always be greater than or equal to 1. $g^{(n)}<1$ is an indicator for nonclassical light and $g^{(n)}=0$ for a perfect single-photon source. However, the relation between sub-Poissonian statistics and anti-bunching requires further care, e.g., see ref.~\cite{zou:90}. Details on the determination of $g^{(n)}$ and measured coincidences can be found in the sections \ref{SI:S2}--\ref{SI:S4} of S.I.
%521
%125
An additional note to add is that the $g^{(n)}$ measured here is a time average of the more general $g^{(n)}(t)$ which is expected to be at a maximum (for classical light) or minimum (for nonclassical light) at time $t=0$ and decay/increase to 1 at $t\geqslant t_c$. Here, $t_c$ is the coherence time of the light source. In this experiment, the measured $g^{(n)}$ will be the average over a SBSL pulse duration which is expected to be in the order of tens of picoseconds. Thus, we expect an even smaller $g^{(n)}$ to be measured if a detector with a shorter timing resolution than that of the SBSL pulse is used. 

Studying photon statistics by employing Photon Number Resolving (PNR) detectors to quantify the statistical properties is an alternative powerful method in quantum optics for characterizing the nature of a light source. For instance, a perfectly coherent light source with constant intensity exhibits Poissonian photon statistics with $g^{(n)}$ = 1, while thermal and nonclassical light sources, on the other hand, demonstrate super-Poissonian $g^{(n)}>1$ and sub-Poissonian $g^{(n)}<1$ photon statistics, respectively~\cite{mandel_wolf_1995, fox2006quantum}. Although experimental measurement schemes can add electronic noise statistics to the inherent photon statistics, this technique can still be employed to classify different light sources based on the origin of their photon emission process~\cite{hlouvsek:19}.
%204

\begin{figure*}[t]
	\centering
	\includegraphics[width=2\columnwidth]{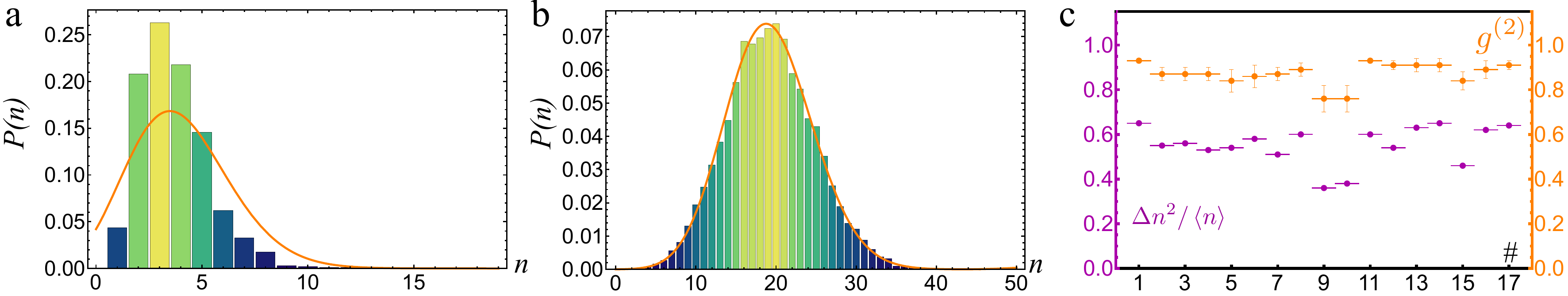}
	\caption{\textbf{Experimental measurement and statistical analysis results from photon number resolving study for SBSL and multimode thermal source}. (a) shows photon-number statistics and the fitted Poisson curve for comparison. $P(n)$ is the probability of finding $n$ photons within the subsegment corresponding to $172.8~\mu$s detection window vs measured photon-number. The measured photon number distribution for SBSL is clearly narrower than the fitted Poisson curve, a clear indication of sub-Poissonian photon number statistics for emitted photons from SBSL. (b) multimode thermal state, photon number distribution of a broadband thermal light source whose spatial profile is first filtered using a single-mode optical fibre. As expected, since the detection window is much longer than the coherence time of the light source, the second-order correlation $g^{(2)}$ is averaged to 1 corresponding to Poissonian statistics, and is clearly seen through the nearly perfect fit of the Poisson curve. (c) shows the measured $g^{(2)}$ and $(\Delta n)^2/\av{n}$ for seventeen different SBSL experiments. Consistently, we observed a $g^{(2)}<1$ and $(\Delta n)^2/\av{n}<1$, when we analysed emitted photons from SBSL experiment.}
	\label{fig:fig3}
\end{figure*}

Photons emitted from a coherent light source with stable intensity and mean photon number $\av{n}$ exhibits Poisson distribution, $P(n)={\av{n}^n}/{n!}\,\, e^{-\av{n}}$, with a photon-number variance $(\Delta n)^2=\av{n}$. While single mode thermal light displays Bose-Einstein distribution with a probability that relates to mean photon number according to $P(n)={\av{n}}^n/{(1+\av{n})}^{n+1}$, and the variance $(\Delta n)^2=\av{n}+\av{n}^2$. However, when multimode thermal light with $N_m$ modes is considered, the variance becomes $(\Delta n)^2=\av{n}+\av{n}^2/N_m$. For many-mode thermal light, i.e. $N_m\gg 1$, the photon-number variance is approximately equal to the mean photon number, $(\Delta n)^2\approx\av{n}$, similar to a Poissonian distribution. On the other hand, sub-Poissonian photon statistics, characterized by $(\Delta n)^2<\av{n}$, cannot be described by classical electrodynamics theory, and is a signature of quantum light sources. Therefore, by comparing the variance in photon number distribution and the mean photon number, one is able to determine whether a light source is classical or quantum in nature. Figure~\ref{fig:fig3}-(\textbf{a}) shows the photon number statistics observed for the SBSL, measured using a PNR camera. When compared to a fitted Poisson curve with the same average photon number, one can clearly observe the photon distribution is sub-Poissonian. Using the relation $g^{(2)}=1+\left((\Delta n)^2-\av{n}\right)/{\av{n}^2}$ , an average $g^{(2)}=0.87\pm0.02$ for SBSL was obtained after performing a series of photon counting measurements. The measured $g^{(2)}$ via photon number resolving detector is close to the previously measured ones, presented earlier with the time-tagging camera. Since the minimum exposure time of the camera is $172.8~\mu$s, which is approximately 5 SBSL pulses, the measured $g^{(2)}$ value should be considered as an upper bound limit. For comparison, in Fig.~\ref{fig:fig3}-(\textbf{b}) the result obtained from a broadband multimode thermal source is presented. The measured photon statistics for the multimode thermal source is clearly Poissonian. See S.I. for more detailed experimental outcomes and statistical analysis results. The photon statistics were measured for several different experiments taken over several weeks. Figure~\ref{fig:fig3}-(\textbf{c}) shows $g^{(2)}$ and $(\Delta n)^2/\av{n}$ for seventeen different SBSL experiments. For all of these experiments, both $g^{(2)}$ and $(\Delta n)^2/\av{n}$ were observed to be below $1$, and thus confirm that the emitted photons from SBSL, unlike a thermal source, is sub-Poissonian. \newline
%329
%

In summary, measurement and characterization of photon number statistics and multichannel time-resolved photon counting were employed to probe the nature of emitted light from SBSL. Results from studying higher-order temporal correlations and direct measurement of the photon number statistics show a clear indication of sub-Poissonian photon statistics, and manifested the quantum nature of the emitted photons. The models based on classical electrodynamics, such as thermal radiation, cannot lead to these observed sub-Poissonian photon statistics --  photons emitted from thermal sources possess super-Poissonian statistics. Results from the measurements suggest photons emitted in sonoluminescence should be originating from a quantum phenomenon. So far, the dynamic Casimir effect (DCE) is the main candidate in quantum theory proposed to describe SBSL~\cite{schwinger1994casimir,eberlein1996theory,milton1997casimir,musha2011another}. However, there have been studies showing that DCE suffers from some major flaws such as requiring the bubble to contract faster than the speed of light~\cite{lambrecht1997comment, unnikrishnan1996comment,garcia1997comment,brevik:99}. We hope that our observation of sub-Poissonian photon statistics will shed some light on the physics of sonoluminescence, though providing a quantum theory to explain our observations will require further investigation and is beyond the scope of this work. Moreover, given the observed strong correlation among generated photons from SBSL, a controlled SBSL would be a good candidate for deterministic quantum photonic sources that is a genuine need in quantum science and technology.
%219
\newline
\newline

\noindent\textbf{Method}\newline
\footnotesize
\noindent\textbf{Higher-order correlation measurements:} Figure~\ref{fig:fig1}-(\textbf{a}) shows the setup schematic view for multiphoton correlation measurements of the SBSL source. SBSL photons from seven different directions are redirected towards the time-tagging camera (TPX3CAM) and imaged onto seven different spots on the camera. The camera has nanosecond time-resolution for each individual pixel and is made single-photon sensitive with the aid of an attached image intensifier (Photonis' Cricket with Hi-QE Red photocathode). This camera, therefore, allows us to record each detected photon's time and pixel coordinates, and consequently, the spatiotemporal correlation among detected photons. The camera and intensifier system together have a timing resolution of approximately 7~ns. Thus, a coincidence gating window of 25 ns was used, and only photons detected within this gating window are considered as coincidence events and used in the correlation calculations. The image intensifier has a spectral sensitivity range between approximately 500\,nm-900\,nm, and we did not use any spectral filters in our setup. Therefore, all photons in this spectral range that are not absorbed by the water (and flask) were detected. Due to the image intensifier converting individual photons into a flash of light, a cluster of pixels on the camera is often illuminated by a single photon event. Thus, each cluster must be regrouped into a single event employing a centroiding algorithm~\cite{ianzano:20}. Also, depending on the light intensity detected on each pixel, each event's measured Time-of-Arrival (ToA) may differ from pixel to pixel even if they are from the same cluster. High photon intensity will cause a steeper signal rise time and, therefore, cross the camera's discrimination threshold earlier than smaller signals, giving an earlier ToA. This time-walk effect also must be corrected in the analysis to improve the camera's timing resolution~\cite{ianzano:20}. After addressing this issue, a timing resolution of approximately 7 ns is achieved.\newline

\noindent\textbf{Photon Number Statistics Measurement:} 
To reveal the underlying nature of sonoluminescence light, we directly investigated the statistics of emitted photons and analysed the probability distribution using a photon number resolving (PNR) camera. For this experiment, we utilized the ORCA-Quest quantitative CMOS (qCMOS) camera from Hamamatsu, which can distinguish the number of arrival photons up to 200 within each pixel at $172.8~\mu$s time resolution. We ensured the photon intensity was far below the photon number limit in all measurements. The camera has 9.4 Megapixels ($4.6~\mu$m$\times4.6~ \mu$m pixel size), and when running in ultra-quiet mode, the RMS readout noise is only 0.27 electrons. The PNR camera has high quantum efficiency, 90\% at $-20^\circ$C around the $475$~nm region, which is also the most intense spectral region of SBSL. Minimizing different sources of noise that may interfere with measuring the intrinsic photon statistic in the experiment is key to unrevealing the fundamental nature of the light source under study. Here, we took 10,000 images with $172.8~\mu$s exposure time and constructed a photon number distribution histogram using the photon number recorded on each pixel. To ensure the authentic creation of actual photon distribution statistics and validate our approach, we compared results obtained from SBSL to that of a stabilized broadband light source with a black-body radiation spectrum (aka thermal source). The variance and mean photon number were measured from the photon number distribution histogram and was compared for 17 different data sets (see S.I.). Probability density of photon detection is plotted against the mean photon number. Considering Mandel's Q-parameter~\cite{Mandel:79} defined as $Q=((\Delta n)^2-\av{n})/{\av{n}}$ and $g^{(2)}=1+Q/\av{n}$ as the figures of merit makes it clear to distinguish between classical ($Q>0$,~ $g^{(2)}>1$) and nonclassical ($Q<0$,~ $g^{(2)}<1$) light.
\newline

\normalsize
\noindent\textbf{Data availability.} The data that support the findings of this study are available from the corresponding author, E.K.(ekarimi@uottawa.ca), upon reasonable request.
%%%
%%%
%%%
\bibliographystyle{unsrt}
%\bibliography{sono-bib}

%%%
%%%
%%%
%%%
\vspace{0.5cm}
\noindent\textbf{Supplementary Information} accompanies this manuscript.
\vspace{1 EM}

\noindent\textbf{Acknowledgments}
\noindent We like to acknowledge the help we received from the Hamamatsu Photonics and the chance to test the ORCA-Quest qCMOS camera. E.K. would like to thank K. Heshami for fruitful discussion and feedback to characterise the quantum features of the SBSL, and R. Fickler and A. Abas for the first attempt to design the sonoluminescence setup. This work was supported by Canada Research Chairs (CRC), Canada First Research Excellence Fund (CFREF) Program, and NRC-uOttawa Joint Centre for Extreme Quantum Photonics (JCEP) via High Throughput and Secure Networks Challenge Program at the National Research Council of Canada. M.R. would like to acknowledge the support of Mitacs Accelerate Industrial Postdoctoral Fellowship.
\vspace{1 EM}

\noindent\textbf{Author Contributions}
E.K: Conceived the idea; M.R., Y.Z., J.H. and E.K.: Designed the experiment; M.R. and Y.Z.: Performed the experiment and collect the data; M.R., Y.Z. and E.K.: Analysed the data; All authors discussed the results and contributed to the text of the manuscript.
\vspace{1 EM}

\noindent\textbf{Author Information}
\noindent The authors declare no competing financial interests. Correspondence and requests for materials should be addressed to E.K. (ekarimi@uottawa.ca).

%%%%%%%
\clearpage
\onecolumngrid
\renewcommand{\figurename}{\textbf{Figure}}
\setcounter{figure}{0} \renewcommand{\thefigure}{\textbf{S{\arabic{figure}}}}
\setcounter{table}{0} \renewcommand{\thetable}{S\arabic{table}}
\setcounter{section}{0} \renewcommand{\thesection}{S\arabic{section}}
\setcounter{equation}{0} \renewcommand{\theequation}{S\arabic{equation}}
\onecolumngrid

\begin{center}
{\Large Supplementary Information for: \\ Observation of Nonclassical Photon Statistics in Single-Bubble Sonoluminescence}
\end{center}
%\appendix
\vspace{1 EM}
%%%%%%%
\section{\label{SI:S1} Experimental setup for generating single-bubble sonoluminescence}
\begin{figure*}[b]
	\centering
	\includegraphics[width=\linewidth]{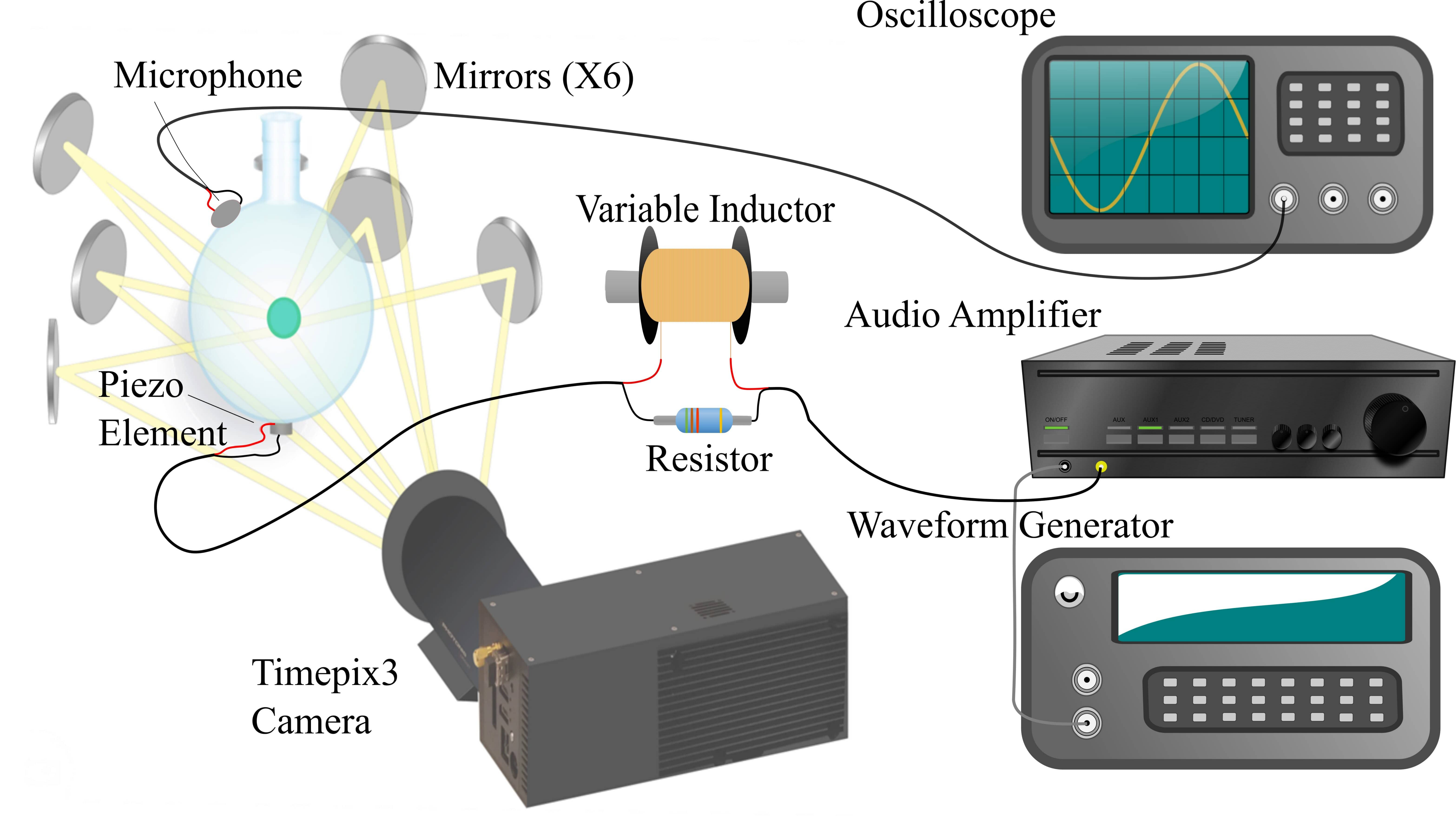}
	\caption{\textbf{Sonoluminescence characterisation experimental setup.} The ~100 ml round flask is filled with degassed water at room temperature. The sound wave at ~28.06 kHz is generated using the function generator, and the audio amplifier intensifies the signal amplitude. The amplified signal is sent to the solenoid and, from there, to the piezoelectric element. We can tune the circuit resonance frequency to match that of the water-filled flask using the movable ferrite rod core in the inductor. Monitoring the microphone signal's amplitude using the oscilloscope can ensure the proper resonance frequency is applied. The time-stamping camera images sonoluminescence light reflected by the six mirrors and the source light. }
	\label{fig:figS1}
\end{figure*}

Different methods are suggested to transfer the ultrasound energy to the liquid to observe sonoluminescence. Various configurations which successfully implemented so far include using an ultrasound horn immersed in the liquid, having two piezoelectric elements glued to the opposite sides of the flask facing each other, and using a single piezo transducer attached to the bottom of the flask, to name a few. In this experiment, a single piezoelectric transducer attached to the bottom of a $~100$~ml round bottom quartz flask ($63$~mm external diameter) using epoxy was used. Another transducer was epoxied to the flask's side to work as the microphone to find the system's resonance frequency. Experiments should be performed in a dark lab and inside a black box to minimize any stray light. Stray lights must be minimized since the sonoluminescence light is faint and therefore not easy for the naked eye or the commercial camera (Canon EOS 550D) to detect. The drive signal is generated using a waveform generator (Agilent 33220A) and amplified in the next stage using an audio amplifier. A successful experiment's main steps include water preparation, signal conditioning, and finding the vessel's containing the liquid medium resonance condition. Although not necessary, only water was used (according to other studies, sulfuric acid makes brighter sonoluminescence light than degassed water) for the sake of simplicity and safety. Water should be free of large particle contamination. In general, the results from qualitative investigations suggest around $~5-10^\circ$C; sonoluminescence light is more intense. Observed sonoluminescence light intensity is a function of temperature since the speed of acoustic waves propagating in the water as well as the amount of dissolved gas are temperature dependent, among other possible reasons. Figure~\ref{fig:figS1} shows the general outline of the experiment and its main components. 
The function (frequency) generator output signal should be stable (frequency and amplitude) around $25-30$~kHz. For the setup used in this experiment, $28.06$~kHz turned out to be the correct resonance frequency. The signal is amplified using an audio amplifier, and sent to an inductor formed from a wire spool ($35$~mH) with a ferrite rod inside. This variable solenoid helps to tune the circuit. From the inductor, the signal is connected to the piezoelectric transducer. It is essential to find the right frequency for trapping bubbles since there is more than one resonance frequency. At the proper resonance frequency, the flask is very sensitive, and exerting force to its surface causes a drastic change in the signal's amplitude from the microphone. Also, a hydrophone can be utilized to read the acoustic field amplitude in the flask (probably more accurate and expensive). One can fine-tune the frequency and adjust the inductance by moving the ferrite rod in and out of the solenoid. When the right frequency is found, the driving signal's amplitude should increase from the audio amplifier. Observing the bubble's behaviour using a commercial camera and reading the microphone amplitude, the amplitude should be increased until the bubble stops jittering and becomes steady. The acoustic field's threshold amplitude is the right amount of sound energy to trap the bubble and see the light without losing it. The flask is mounted on a translational stage connected to a 3-dimensional stage for fine-tuning the light source for further quantum optics experiments. 
\begin{figure*}[hbt]
	\centering
	\includegraphics[width=\linewidth]{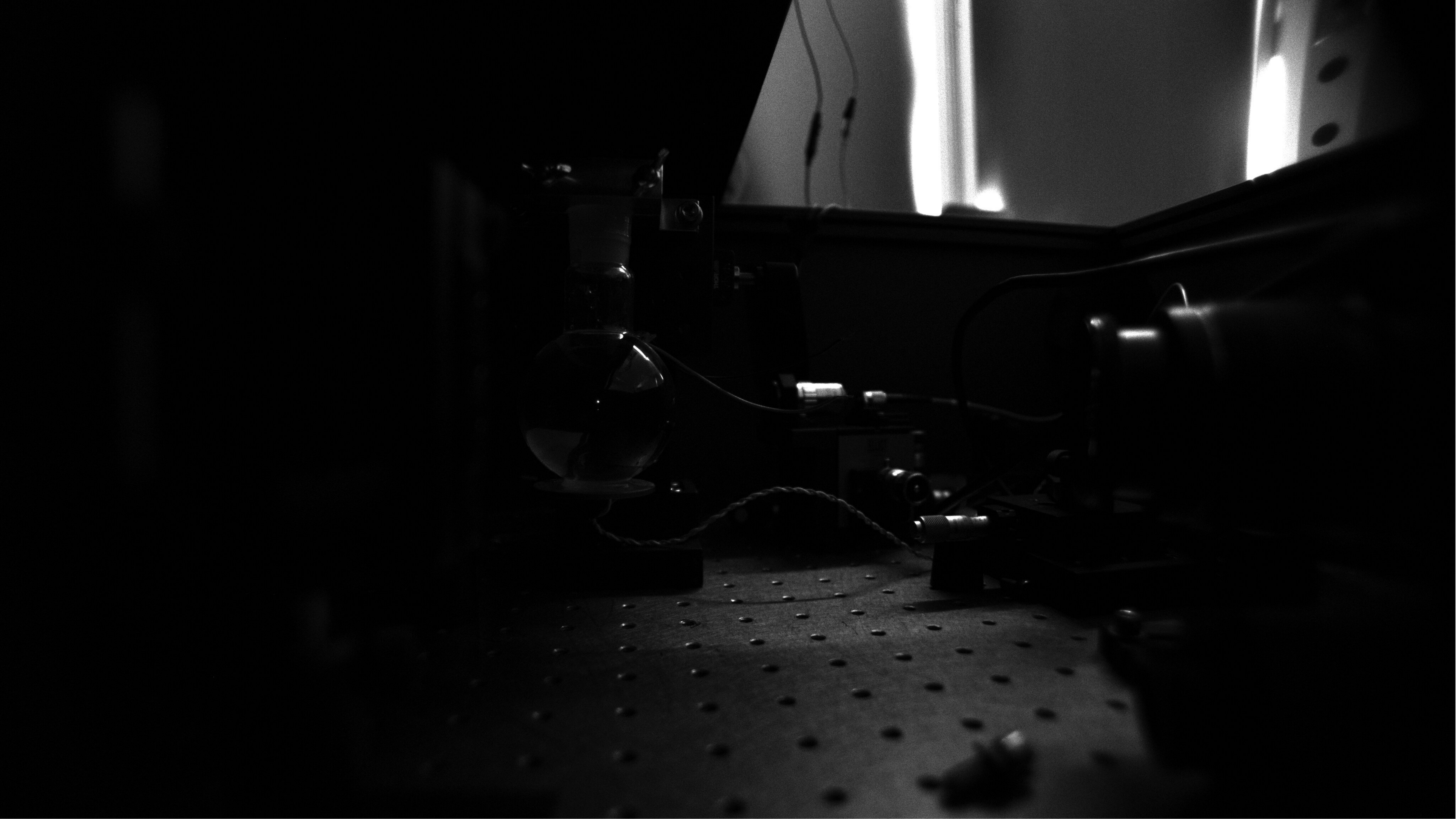}
	\caption{\textbf{Sonoluminescence light in degassed water.} The small dot approximately at the centre of the flask is the sonoluminescence light, visible to the unaided eye in the dark. Photo is taken using the ORCA-Quest camera 5~s exposure time. Note that during data taking, the enclosure were sealed properly to avoid any stray light entering to the box.}
	\label{fig:figS2}
\end{figure*}
Figure~\ref{fig:figS2} is a picture of the sonoluminescence round flask with the glowing bubble in the middle taken ORCA-Quest camera with 5~s exposure time. The light appears as a small faint dot, almost located in the middle of the flask. In a completely dark room, the light is visible with the naked eye.

%%%%%%%
\section{\label{SI:S2}Second-order correlation function measured with single-photon avalanche photodiodes}
The initial second-order correlation function $g^{(2)}$ measurements were performed using two single-photon avalanche photodiodes ($\tau$-SPAD-50) between two channels ($s_3$-the source and $s_4$). The emitted photons from the SBSL in two channels $s_3$ and $s_4$ are coupled into multimode optical fibres utilizing two $10\times$ microscope objectives, and then detected by two single-photon avalanche photodiodes (APD). The APDs quantum efficiency at 550~nm is about 70\% with a dark count and deadtime of $50$ counts per second and $50$~ns, respectively. A time tagging unit (Swabian instruments \qo{Time Tagger 20}) is used to measure the coincidences between the two APDs. Results of this coincidence measurement showed strong time-correlation between detected photons, a clear indication of photon pair generation. The measured FWHM is ~5~ns. The measured coincidences and $g^{(2)}$ (see S5 for the calculation of $g^{(2)}$) is shown in Table~\ref{tab:tabs1}. This measurement was a motivation to further explore the higher-order correlations function by switching to a time-tagging camera.
\begin{table}[tb]
  \centering
 \caption{\label{tab:tabs1} \textbf{Coincidences and $g^{(2)}$ measured for SBSL with multimode fibers and APDs.} Background in the Singles 1 and 2 are 56 counts/s and 48 counts/s, respectively.}
\begin{tabular*}{0.8\textwidth}{@{\extracolsep{\fill} }  c|ccccccc}
    \hline\hline
              &     Set 1   & Set 2  &   Set 3  &  Set 4   & Set 5  &   Set 6  &  Set 7  \\ \hline\hline
   Integration time (s)    &  250 &	150 &	120	& 100 &	100 &	200 &	200    \\ \hline
Singles 1 (background subtracted) &	81,041 &	54,059 &	52,713 & 46,242 & 52,491 & 107,514 & 61,378 \\ \hline
Singles 2 (background subtracted) &	74,592 &	48,677 &	44,796 &	36,487 &	42,849 &	88,333 &	84,273 \\ \hline
Measured Coincidences &	748 &	509 &	634	& 517 &	708	& 1538 & 768  \\ \hline
$g^{(2)}$	& 0.87 & 	0.82 &	0.91 &	0.87 &	0.89 &	0.91 &	0.84 \\ \hline \hline
\end{tabular*}
\end{table}

\section{\label{SI:S3}Higher-order correlation function measured with time-tagging camera}
A time-tagging camera (TPX3CAM) with an attached image intensifier (Photonis' Cricket with Hi-QE Red photocathode) was used for the time-correlation setup for measuring higher orders of correlation between the SBSL photons. The quantum efficiency of the image intensifier was measured to be approximately 5\% on average in the 500-900\,nm wavelength region and the camera offers the time-stamping capability for individual pixels ($256\times256$ pixels with $55\times55~\mu\,m^2$ pixel size) of the camera's active detection area. These features made this experiment feasible and offered a new approach to characterise sonoluminescence light. Though the camera inherently has a 1.5 ns time resolution, however, after clustering and time-over-threshold (TOT) corrections, as discussed in the Methods section of the main test, a timing resolution of approximately 7\,ns (FWHM) is achieved. Intensified camera integrated with the intensifier can detect and time-tag approximately 10\,M photons per second across all its pixels. Time correlations were recorded by using six mirrors around the flask to reflect the sonoluminescence light towards the camera, overall, seven photon spots on the camera. Along with the source spot, nearly at the centre of the flask, this configuration can simultaneously record the position and arrival time of photons at seven different points (seven light detection channels).  The current setup can be employed for multichannel photon correlation measurements. Although this camera has been designed and built with high energy and particle physics applications such as imaging ions in mind, it proved extremely useful in quantum optics experiments.

\section{\label{SI:S4}Multiphoton time-correlated photon counting}
In order to characterise the state of light, we measured the second-order intensity correlation function, $g^{(2)}$, as well as higher orders up to five to investigate the nature of the emitted light.  Statistical measurements and calculating temporal intensity correlation functions of two-or more photon channels can reveal a lot of information regarding the nature of the field. The second-order correlation function, also known as the degree of second-order coherence, $g^{(2)} (\tau)$ can be given in terms of the intensity $I(t)$ as
\begin{align}
g^{2}(\tau)=\frac{\av{I(t)I(t+\tau)}}{\av{I(t)}^2}\,
\end{align}
where $I(t)$ is also proportional to the photon number. It is extremely important when it comes to characterising the statistical properties of the emission source since it is a reliable indicator of the quantum or classical nature of light source. 

The n-th order correlation (normalized Glauber's correlation functions) function is defined as,
\begin{align}
g^{n}(t_1,t_2,\ldots,t_n)=\frac{\av{I_1(t_1)I_2(t_2)\ldots I_n(t_n)}}{\av{I_1(t_1)}\av{I_2(t_2)}\ldots\av{I_n(t_n)}}\,
\end{align}
since there is no position dependency in this experiment, one can ignore the position dependency and the correlation function will be merely a function of time. For this experiment, since the detector timing resolution is longer than the SBSL pulse duration, our measured $g^{(n)}$ is time averaged over the entire pulse. In terms of the physically measured parameters, the single channel and coincidence counts, the 2nd order correlations are calculated according to
\begin{align}
g^{2}_{(i,j)}=\frac{C_{(i,j)}\,R\,T}{S_i\,S_j}\,
\end{align}
where $C_{(i,j)}$ is the total coincidence counts within the data recording time $T$ between channels $(i,j)$, $S_i$ and $S_j$ are the single channels counts (background subtracted) within time $T$ , $R=28,060$~Hz is the SBSL pulse repetition rate and $T=22$~s.

For n-th order correlations, this is given by:
\begin{align}
g^{(n)}_{(i,j)}=\frac{C_{(n)}\left(R\,T\right)^{n-1}}{\Pi_{i=1}^{n} S_i}\,
\end{align}
where $C_{(n)}$ is the $n$-fold coincidence between channels $(s_i...s_n)$. The number of background subtracted photon detection events over $22$ seconds for the channels/spots from $s_1$ to $s_7$ are $124,627$; $119,499$; $243,295$; $109,486$; $111,898$; $77,269$ and $75,708$, respectively. Here, only events within a 10-pixel radius around the centre of each bright spot were taken into account; an average background of $4.3$ photons per second per pixel was measured, an equivalent of $30,678$ photons in $22$ seconds within the $10$-pixel radius circle. The result of coincidence counts for all the possible permutations of channels are summarised in Fig. S4. Using the in-house developed code for analysing the data, coincidence is calculated for all the possible permutations between different channels. Panels include different possible orders of permutation of channels. \newline

Although using avalanche photodiodes offers detection with higher time-resolution, employing the time-tagging camera helps to reduce the experimental complexity of multi-channel coincidence measurements. Although it is simple to increase the number of spots/channels, a much longer data acquisition time will be required for ($n>5$)-channel coincidence detections, subjecting the measurement to instabilities in the SBSL source.%
\begin{figure*}[hbt]
	\centering
	\includegraphics[width=\linewidth]{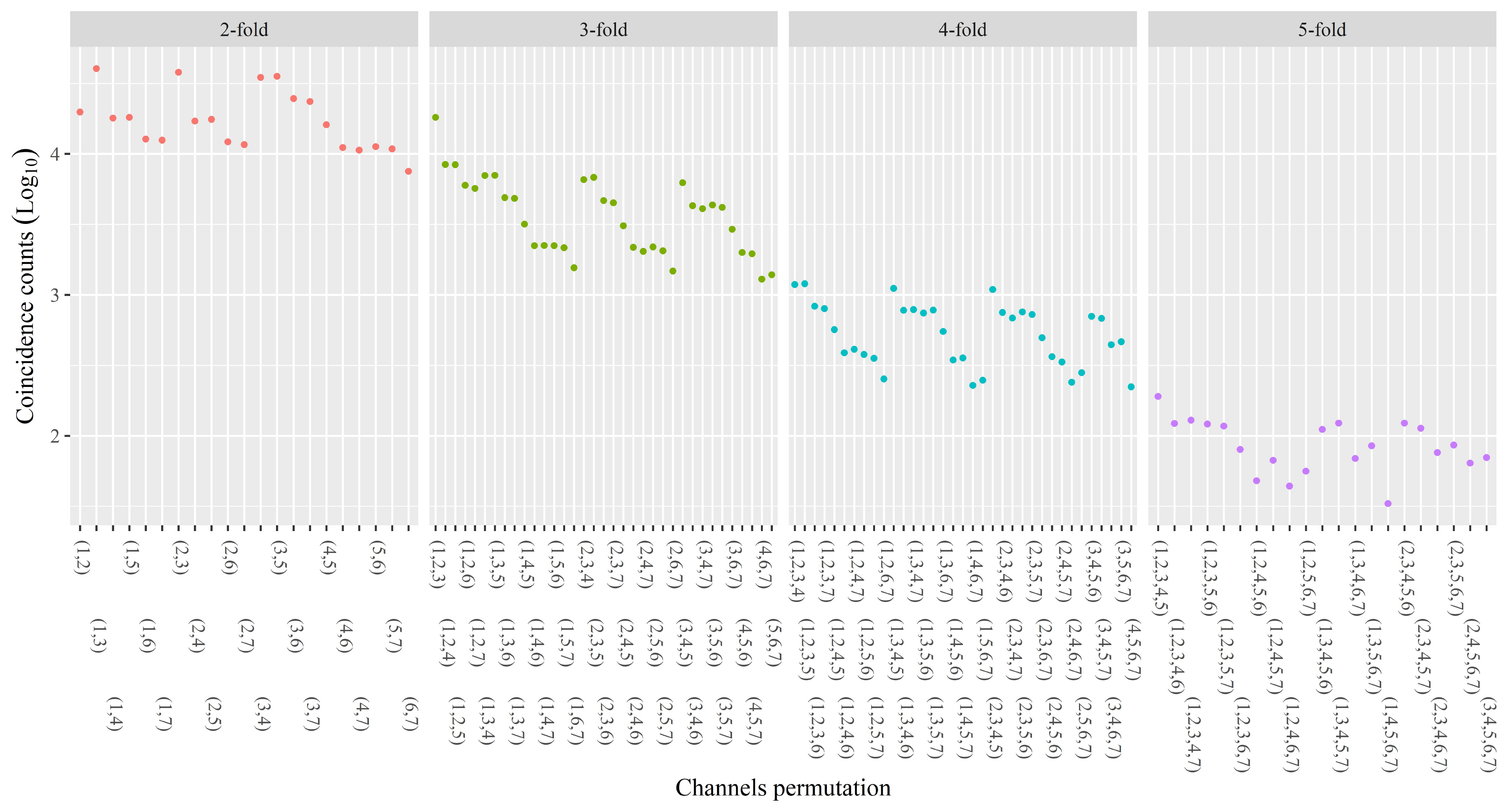}
	\caption{\textbf{Summary of all measured coincidences for all possible permutations of the seven channels (e.g., (1,2); (1,3); (1,3,4); etc.) for multi-fold coincidences.} Combination of all coincidence counts for different permutation on all seven channels, possible for n-fold coincidence measurements for n=2-5. The total integration time is 22 seconds with a 25 ns gate width. Only events within a 10-pixel radius around the centre of each bright spot were taken into account in the analysis. The exponential decay of the coincidence counts by increasing the number of folds is a characteristic that can be attributed to the camera's quantum efficiency. }
	\label{fig:figS4}
\end{figure*}

\section{Photon number resolving measurement}
We recorded an area containing $100\times100$ pixels on the PNR camera with the SBSL spot  at around 10 pixels in diameter. Every pixel shows a detection of a few photons per frame, well within the range of the camera specifications. 

Variance and mean photon numbers were acquired from statistical analysis and compared for 17 data sets to determine the inherent nature of photon statistics. Considering Mandel $Q$ parameter defined as $Q=\left((\Delta n)^2-\av{n}\right)/\av{n}=\av{n}(g^{2}-1)$, makes it clear to distinguish between classical and nonclassical light. Negative $Q$ is an indication of nonclassical light and only possible if the statistics is sub-Poissonian. For simplicity, we also introduced another figure of merit $E={(\Delta n)^2}/{\av{n}}$ and obviously an $E<1$ is a clear sign of sub-Poissonian photon statistics. To calculate the second-order correlation function:

\begin{align}
g^{(n)}=\frac{\av{\hat{a}^\dagger(t_1)\ldots\hat{a}^\dagger(t_n)~\hat{a}(t_n)\ldots\hat{a}(t_1)}}{\av{\hat{a}^\dagger(t_1) ~\hat{a}(t_1)}\ldots\av{\hat{a}^\dagger(t_n) ~\hat{a}(t_n)}}\,
\end{align}
we employed $g^{(2)}=1+\left((\Delta n)^2-\av{n}\right)/\av{n}^2$  which by comparing the results obtained from a stabilized thermal source, it does produce reliable results for this experiment. Table~\ref{tab:tabs2} shows the summary of results from 17 measurements and the statistical analysis using the maximum-likelihood approach performed subsequently. 

\begin{table}[tb]
  \centering
 \caption{\label{tab:tabs2} \textbf{Summary of statistical analysis of recorded photons using a photon number resolving camera for 17 different data runs.} STDEV is the standard deviation and for each histogram, we recorded $10,000$ sequential images with $172.8~\mu$s exposure time. Only one pixel with highest intensity was considered in this analysis.}
\begin{tabular*}{0.8\textwidth}{@{\extracolsep{\fill} }  c|ccccccc}
    \hline\hline
Sample \#  & $\av{n}$ & $(\Delta n)^2$ & $g^{(2)}$ & $Q$ & $E$ & STDEV & $g^{(2)}$ uncertainty \\ \hline\hline
1&5.36&3.49&0.93&-0.35&0.65&1.87&0.01 \\ \hline
2&3.51&1.95&0.87&-0.45&0.55&1.39&0.03 \\ \hline
3&3.50&1.96&0.87&-0.45&0.56&1.40&0.03 \\ \hline
4&3.52&1.88&0.87&-0.47&0.53&1.37&0.03 \\ \hline
5&2.95&1.60&0.84&-0.46&0.54&1.26&0.05 \\ \hline
6&2.90&1.68&0.86&-0.42&0.58&1.30&0.05 \\ \hline
7&3.76&1.93&0.87&-0.49&0.51&1.39&0.03 \\ \hline
8&3.54&2.12&0.89&-0.40&0.60&1.46&0.03 \\ \hline
9&2.64&0.95&0.76&-0.64&0.36&0.98&0.06 \\ \hline
10&2.61&0.99&0.76&-0.62&0.38&0.99&0.06 \\ \hline
11&6.00&3.62&0.93&-0.40&0.60&1.90&0.01 \\ \hline
12&5.17&2.78&0.91&-0.46&0.54&1.67&0.02 \\ \hline
13&3.99&2.51&0.91&-0.37&0.63&1.58&0.03 \\ \hline
14&3.80&2.47&0.91&-0.35&0.65&1.57&0.03 \\ \hline
15&3.42&1.56&0.84&-0.54&0.46&1.25&0.04 \\ \hline
16&3.42&2.11&0.89&-0.38&0.62&1.45&0.04 \\ \hline
17&4.15&2.67&0.91&-0.36&0.64&1.64&0.02 \\ \hline \hline
\end{tabular*}
\end{table}

\end{document}